# Density Functional Modeling and Total Scattering Analysis of the Atomic Structure of a Quaternary CaO-MgO-Al$_2$O$_3$-SiO$_2$ (CMAS) Glass: Uncovering the Local Environment of Magnesium

[1]Kai Gong, [1,2]V. Ongun Özçelik, [1]Kengran Yang, [1]Claire E. White

[1]Department of Civil and Environmental Engineering, Andlinger Center for Energy and the Environment, Princeton University
[2]Department of Chemistry and Biochemistry, Materials Science and Engineering Program, University of California San Diego (current address)

**Abstract**

Quaternary CaO-MgO-Al$_2$O$_3$-SiO$_2$ (CMAS) glasses are important constituents of the Earth's lower crust and mantle, and they also have important industrial applications such as in metallurgical processes, concrete production and emerging low-CO$_2$ cement technologies. In particular, these applications rely heavily on the composition-structure-reactivity relationships for CMAS glasses, which are not yet well established. In this study, we developed a robust method that combines force-field molecular dynamics (MD) simulations and density functional theory (DFT) calculations with X-ray/neutron scattering experiments to resolve the atomic structure of a CMAS glass. The final structural representation generated using this method is not only thermodynamically favorable (according to DFT calculations) but also agrees with experiments (including X-ray/neutron scattering data as well as literature data). Detailed analysis of the final structure (including partial pair distribution functions, coordination number, oxygen environment) enabled existing discrepancies in the literature to be reconciled and has revealed new structural information on the CMAS glass, specifically, (i) the unambiguous assignment of medium-range atomic ordering, (ii) the preferential role of Ca atoms as charge compensators and Mg atoms as network modifiers, (iii) the proximity of Mg atoms to free oxygen sites, and (iv) clustering of Mg atoms. Overall, this new structural information will enhance our mechanistic understanding on CMAS glass dissolution behavior, including dissolution-related mechanisms occurring during the formation of low-CO$_2$ cements.

# 1    Introduction

The structure and properties of silicate glasses are of significant interest to many scientific and technological fields such as condensed matter physics, geology, glass science, materials chemistry, energy, medicine and advanced communication systems.[1, 2] In particular, CaO-Al$_2$O$_3$-SiO$_2$ (CAS) ternary glasses are one of the most studied glass systems due to their advantageous optical, mechanical and chemical properties,[3-5] rendering them an attractive option for a wide range of applications such as nuclear waste encapsulation, high performance glasses, ceramics, metallurgical process, and cements.[6] The structure of a CAS glass generally consists of silicate and aluminate tetrahedra (commonly referred as network formers) connected via their bridging oxygen (BO) atoms to form a network, which is modified by calcium cations (network modifiers). The impact of calcium on the aluminosilicate network structure is two-fold: (i) to charge-balance the negative charge associated with aluminate tetrahedra (i.e., AlO$_2^-$), and (ii) to break the aluminosilicate network creating non-bridging oxygen (NBO) atoms. The introduction of network modifiers (e.g., Ca) alters the structural properties of aluminosilicate glasses (e.g., relative amounts of BO and NBO) and consequently changes their physical, optical, mechanical, thermo and chemical properties.[7-9]  Hence, the structural properties of ternary CAS glasses have been widely studied both from an experimental[10-25] and computational[7, 26-33] viewpoint.

Magnesium is another common network modifier that has an impact on the aluminosilicate network structure similar to calcium.[34] In fact, quaternary CaO-MgO-Al$_2$O$_3$-SiO$_2$ (CMAS) glasses are important constituents of the Earth's lower crust and mantle[35, 36] and have industrial applications including metallurgical processes, concrete production and emerging low-CO$_2$ cement technologies.[37-44] For instance, both CMAS (e.g., blast-furnace slag from steel manufacturing process) and CAS (e.g., coal-derived class C fly ash) glasses are often used to partially replace ordinary Portland cement (OPC) in concrete production to (i) enhance the mechanical properties and long-term durability of concrete and (ii) lower the CO$_2$ emissions associated with use of OPC.[44] In addition, both CMAS slag and CAS fly ash are important precursor materials for synthesis of alkali-activated materials (AAMs), which constitute a class of low-CO$_2$ cements with excellent mechanical, thermal and chemical properties when properly formulated.[43] Both applications have great potential to significantly reduce the environmental impact of the current cement industry, which accounts for 8-9% of global anthropogenic CO$_2$ emissions.[45] Furthermore, CMAS glass has been identified as a major source of corrosion and premature failure for ceramic thermal barrier coatings used to enhance the high-temperature behavior of alloys in spacecraft and aircraft.[46, 47]

Despite the importance of CMAS melts and glasses in industrial applications and the associated environmental benefits, research on the composition-structure relationships of these glasses is limited compared to CAS glasses. However, these relationships are important to obtain since the structural properties of aluminosilicate glasses control, to a large extent, their chemical reactivity in aqueous environments.[48, 49] It is also known that the attributes of CAS/CMAS glasses in blast-furnace slag influence the strength development, pore structure evolution and long-term durability of OPC and AAM systems.[37, 39, 40, 43, 44, 50, 51] For instance, it has been reported in several studies that the Mg content of CMAS glasses has a large impact on the properties and performance of the resulting AAM.[39, 52, 53] Douglas et al. showed that increasing the MgO content of CMAS glass from 9 to 18 wt. % led to tripling of the 28-day compressive strength of the resulting AAMs.[52] It has also been shown that AAMs based on Mg-rich CMAS glass exhibit superior resistance to carbonation as compared with AAMs based on low-Mg CMAS glass.[41, 53] To fully harness the benefits of CMAS glasses in these applications, it is critical to develop the composition-structure relationships for the CMAS glass systems, and this necessitates the development of detailed atomic structural representations.

Computational tools such as *ab initio* and force-field molecular dynamics (MD) simulations have been used to predict glass structures, uncovering important structural details that are difficult to obtain solely with experiments. Specifically, force-field MD simulations have been widely used to predict the structure and properties of various silicate glasses and melts, including CAS[7, 27, 28, 30-33, 54] and CMAS[38, 55-59] glass systems. A key advantage of force-field MD simulations compared with those based on *ab initio* MD is their relatively high computational efficiency, however, the accuracy of these simulations is highly dependent on the accuracy of the chosen force-field for the material in question[31], where the force-field is developed typically by refining the force-field parameters against limited experimental data and/or *ab initio* calculations.[60] Alternatively, a glass structure can be generated using *ab initio* MD in a more accurate and unbiased manner, where the electronic structure calculations based on the Schrödinger equation are used instead of force-fields. However, the major drawback of *ab initio* MD is its high computational demand that limits its application only to relatively small systems (often less than 100 atoms)[26, 29, 61]. Studies have shown that modeling of silicate glass structures based on small systems (e.g., ~100 atoms) exhibit strong finite size effects on the structural properties (e.g., radial distribution functions and bond-angle distributions).[26, 27]

In this study, we first present a robust protocol that harnessed the benefits of both force-field MD simulations and density functional theory (DFT) calculations to generate ten realistic structural representations for a quaternary CMAS glass with ~440 atoms. This protocol involved subjecting five randomly generated structures to a melt-quench process using force-field MD simulations (widely used for modeling of silicate glass structure [38, 62, 63]) to obtain ten amorphous structural representation for the CMAS glass. These structures subsequently underwent DFT geometry optimization calculations to further improve the accuracy of the structural representations. Furthermore, one DFT-optimized structure was subjected to an interactive process alternating between reverse Monte Carlo (RMC) refinement (where the atom positions were refined against X-ray and neutron total scattering data) and DFT geometry optimization (where the chemical implausibility generated during RMC refinement was addressed) to assess the need for guidance from experimental data to obtain an accurate structural representation. Based on this protocol and analysis of the resulting ten final structural representations, we present new structural information on the CMAS glass and compare our results with literature data. Specifically, key attributes that are computed and compared include the partial pair distribution functions, coordination numbers, oxygen environments and distribution of the network modifiers around oxygen species. Overall, this study highlights the power of combining force-field MD simulations and DFT calculations to generate realistic structural representations for a CMAS glass; the method should be readily transferable to other glass systems and related amorphous materials, and will be particularly helpful for studying dissolution kinetics and mechanisms of glasses in aqueous environments when combined with experimental techniques such as *in situ* pair distribution function (PDF) analysis.

## 2 METHODS

### 2.1 Experimental Details

A quaternary CMAS glass powder with a chemical composition of 42.3 wt. % CaO, 32.3 wt. % $SiO_2$, 13.3 wt. % $Al_2O_3$, and 5.2 wt. % MgO (the Australia slag in ref. [37], other minor oxides not reported here nor included in the simulations) is used in this investigation. This CMAS glass has a similar chemical compositions to a glass structure reported in the literature that was produced using force-field MD.[59] Although the sample contains a small amount of other oxides (the largest being sulfate, $SO_3$, at 2.9 wt. %), their contribution to the X-ray and neutron data are minimal, and therefore only the CMAS glass structure is refined against the experimental data.

X-ray total scattering data were collected on the sample at room temperature on the 11-ID-B beam line at the Advanced Photon Source, Argonne National Laboratory, using a wavelength of 0.2114 Å and a Perkin-Elmer amorphous silicon two-dimensional image plate detector.[64] The wavelength was selected to provide a compromise between high flux (statistics), Q-resolution, and a sufficient maximum momentum transfer. The sample was measured in a 1 mm diameter polyimide capillary. The program Fit2D[65, 66] was used to convert data from 2D to 1D with $CeO_2$ as the calibration material. The pair distribution function (PDF), $G(r)$, is calculated by taking a sine Fourier transform of the measured total scattering function $S(Q)$, where $Q$ is the momentum transfer, as outlined by Egami and Billinge.[67] The X-ray PDF data were obtained using PDFgetX2,[68] with a $Q_{max}$ of 20 Å$^{-1}$. The instrument parameters ($Q_{broad} = 0.016$ Å$^{-1}$ and $Q_{damp} = 0.035$ Å$^{-1}$) were refined by using the calibration material ($CeO_2$) and the refinement program PDFgui.[69]

Neutron total scattering data were collected on the NPDF instrument at the Lujan Neutron Scattering Center, Los Alamos National Laboratory.[70] The sample was loaded in a vanadium can and measured for 8 hrs at room temperature. Standard data reduction for generation of the neutron PDF was performed using the PDFgetN software,[71] including a background subtraction to remove incoherent scattering.[72] A $Q_{max}$ value of 20 Å$^{-1}$ was used to produce the PDF. The neutron instrument parameters were produced using a silicon calibration material ($Q_{broad} = 0.00201$ Å$^{-1}$ and $Q_{damp} = 0.00623$ Å$^{-1}$).

## 2.2    Computational Methods

To generate detailed structural representations for the CMAS glass measured above, we performed force-field MD simulations followed by DFT geometry optimization on a periodic box consisting of 439 atoms. Furthermore, given that previous investigations on amorphous materials have shown that, in addition to such computational methods, refinement against experiment data at some stage during the process is required to obtain accurate structural representations,[73-76] in this study such a refinement has been carried out and its influence on the resulting structure has been assessed. Details on the refinement are outline at the end of this section.

All force-field MD simulations were performed with the ATK-Forcefield module in the Virtual NanoLab (VNL) software package.[77, 78] First, five random structures consisting of 439 atoms each with a chemical composition of $(CaO)_{82}(MgO)_{14}(Al_2O_3)_{14}(SiO_2)_{59}$ (similar chemical composition as the experimental sample discussed above) were generated in cubic unit cells by using the amorphous prebuilder provided in VNL. The size of the cell was selected based on two competing considerations: (i) a minimum of ~200 atoms are required to limit the artificial finite size effects on the structural properties of CAS glasses,[27] and (ii) the prohibitive computational demand of a large system size for subsequent DFT calculations. The density of the unit cell was initially set at 2.40 g/cm$^3$, which is the estimated density for the CMAS glass at a temperature of 5000 K (detailed calculations for this density estimate and justification of the approach are given in the Supporting Information).[79] For all force-field MD simulations, the interatomic potential and parameters developed by Matsui for crystals and melts of the CaO-MgO-Al$_2$O$_3$-SiO$_2$ system were used.[80]

As briefly outlined in Figure 1, each random structure was first subjected to an MD simulation at 5000 K for 1 ns to ensure the loss of the memory of the initial configuration and to reach an equilibrated melt state. The melt was then quenched using MD from 5000 to 2000 K in 3 ns followed by equilibration at 2000 K for 1 ns, before being further quenched from 2000 to 300 K in 3 ns, followed with another 1 ns equilibration time at 300 K. The MD cooling rates of 1.0 and 0.57 K/ps were used here because it has been shown for silicate glasses that the structural properties of the resulting glasses (especially short-range structural ordering, such as the pair distribution functions, bond angles and coordination numbers) are close to convergence at MD cooling rates lower than 1 K/ps.[62, 63] The canonical (NVT with the Nosé Hoover thermostat) ensemble and a time step of 1 fs were used for all the MD simulation steps above, while the density of the unit cell volume was adjusted to numerically estimated values (calculations shown in the Supporting Information) at the start of each equilibration step as shown in Figure 1b. The density of the 300 K final MD structures (2.87 g/cm$^3$) agrees with experimental data on CMAS glasses that have similar compositions.[39, 40] The evolution of ground-state energy of one 300 K MD structure (using single point DFT energy calculation) as a function of cell volume (Figure S1 in the supporting information) further confirms that the estimated density is accurate. Two configurations during the last 500 ps of the MD equilibration step at 300 K (separated by 500 ps) were extracted, leading to a total of ten structures for subsequent DFT calculations.

The configurations extracted from the MD simulations were then subjected to DFT geometry optimizations using the VASP software (version 5.4.1).[81] The purpose of the DFT calculations was to further improve the chemical feasibility of the MD-generated structures. All DFT calculations were performed with the GGA-PBE exchange-correlation functional (using PAW potentials) where the Brillouin zone was sampled using a $2 \times 2 \times 2$ Monkhorst-Pack mesh for k-points. Atomic positions were optimized using the conjugate gradient method, where the total energy was minimized with the cell density fixed at 2.87 g/cm$^3$. For the geometry optimization, a "low" precision was initially employed, where an energy convergence criterion of $10^{-2}$ eV (i.e., EDIFF in INCAR file) was adopted and a relatively large level of Gaussian smearing (0.2 eV width of smearing) was employed to aid convergence. The structure was further optimized using "low" precision without smearing before being subjected to another round of geometry optimization using the "accurate" setting, where the energy convergence criterion was $10^{-3}$ eV.

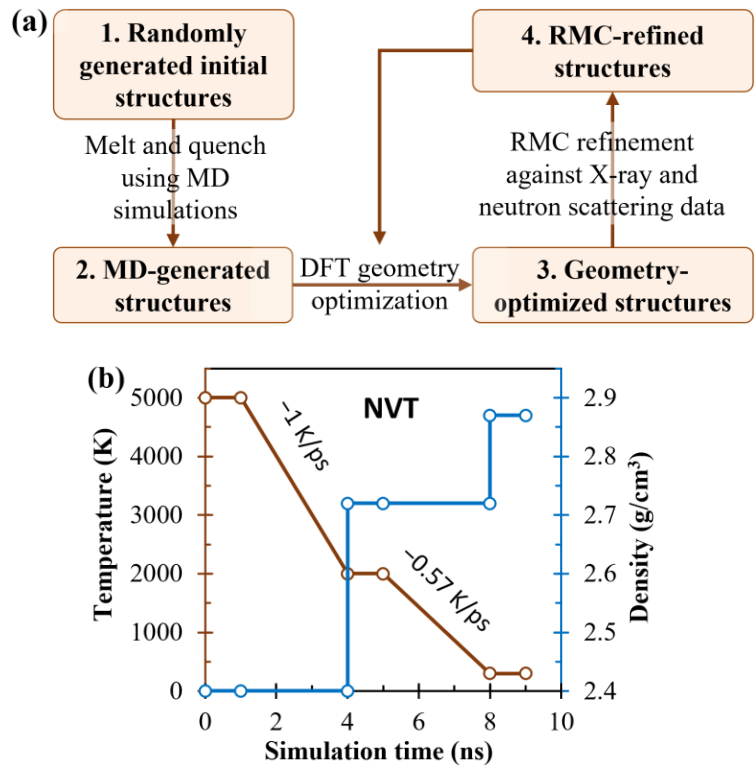

Figure 1. Schematic view of (a) the overall methodology used to generate accurate structural representations of the CMAS glass, and (b) details on the individual MD simulation steps.

In addition to the MD simulations and DFT calculations used to obtain the ten structural representations of the CMAS glass, one of these structures was subjected to an iterative process alternating between (i) an RMC refinement where all the atomic positions were refined against X-ray and neutron total scattering data and (ii) a DFT geometry optimization (loop involving states 3 and 4 in Figure 1a). The purpose of this iterative process is to enable the structure to more widely explore the potential energy surface of the system, and potentially find similar or lower local minima on the potential energy surface with improved agreement with the experimental data. Similar iterative processes have been successfully utilized to generate structural representations for amorphous magnesium carbonate[73], metakaolin[74], and silicon[75, 76].

RMC refinements were performed using RMCProfile,[82] where all atoms were allowed undergo move (max. displacement of 0.5 Å) and swap events whilst being subjected to the physical constraints outlined in Table 1. The probability of swapping for different atom-atom pairs as seen in Table 1 was set at 0.1. The level of agreement, $\chi^2$, is defined in equation 1, where $G_j^{exp}$ and $G_j^{calc}$ are the experimental and calculated values for point $j$ of the PDF, respectively.

$$\chi^2 = \sum_i \chi_i^2 = \sum_i \left( \frac{\sum_j \left( G_j^{exp} - G_j^{calc} \right)^2}{\sigma_i^2} \right) \tag{1}$$

$\sigma_i$ in equation 1 is a weighting factor assigned to dataset $i$, with a smaller weighting factor giving a larger contribution to $\chi^2$ for the corresponding dataset. Since the X-ray dataset contains more structural features between ~3 and 10 Å than the neutron dataset, the X-ray data is more heavily weighted during the RMC refinement process by assigning a smaller $\sigma_i$ value to the X-ray PDF (0.1, as opposed to 0.3 for the neutron data). The move and swap events were accepted or rejected with a given probability depending on whether the individual event led to an improved fit with the X-ray and neutron PDF data, as measured via the change in $\chi^2$ ($\Delta\chi^2$). Specifically, all events that led to an improved fit with the experimental data were accepted (negative $\Delta\chi^2$ value), while the probability of acceptance of those that led to a worse fit (positive $\Delta\chi^2$ value) is determined using equation 2.

$$P = \exp\left(-\frac{\Delta\chi^2}{2}\right) \tag{2}$$

The RMC refinement was considered to have converged after ~800,000 iterations where the percentage of accepted events decreased below 1%. Different weighting factors were used for the different physical constraints (as shown in Table 1), because certain coordination states (e.g., 4-fold Si) are more stringent than others (e.g., 4-fold Al).

Table 1. Parameters used for the RMC refinement, including minimum interatomic distances between specific atom-atom pairs, weighting factors for the datasets and coordination constraints, and swap probabilities for specific atom-atom pairs.

| Atom-atom pair | Minimum interatomic distance (Å) | Dataset | Weighting factors |
|---|---|---|---|
| Mg-Mg | 2.3 | X-rays | 0.1 |
| Mg-Ca | 2.5 | Neutron | 0.3 |
| Mg-Si | 2.45 | | |
| Mg-Al | 2.4 | **Constraint** | **Weighting factor** |
| Mg-O | 1.8 | IV-fold Si constraint (cutoff distance of 2.2 Å) | 0.001 |
| Ca-Ca | 2.5 | | |
| Ca-Si | 2.7 | IV-fold Al constraint (cutoff distance of 2.5 Å) | 0.9 |
| Ca-Al | 2.5 | | |
| Ca-O | 2.15 | **Atom-atom pair** | **Swap probability** |
| Si-Si | 2.8 | Mg-Ca | 0.1 |
| Si-Al | 2.8 | Mg-Si | 0.1 |
| Si-O | 1.5 | Mg-Al | 0.1 |
| Al-Al | 2.7 | Ca-Si | 0.1 |
| Al-O | 1.6 | Ca-Al | 0.1 |
| O-O | 2.4 | Si-Al | 0.1 |

The PDFs (both X-ray and neutron) of the final structural representations were produced using the PDFgui software.[69] The atomic displacement parameters were set at $u_{ii}^2 = 0.003$ Å$^2$, and the experimentally determined values for the $Q$-dependent instrument resolution ($Q_{damp}$) and peak broadening ($Q_{broad}$) parameters were used. The level of agreement between simulated and experimental PDFs was assessed in terms of the $R_w$ value as defined in the PDFgui software,[69] where a smaller $R_w$ value implies better agreement.

## 3    RESULTS AND DISCUSSION

### 3.1    Experimental X-ray and Neutron Data

The experimental X-ray total scattering data for the CMAS glass powder are displayed in Figure 2a, which shows that this glass sample is predominately amorphous, as evidenced by the absence of any obvious Bragg peaks. The neutron total scattering data in Figure 2a, which were collected at a much higher $Q$ resolution than the X-ray data, do exhibit several small Bragg peaks indicative of a very small crystalline impurity. However, the contribution of the crystalline phase(s) to the atom-atom correlations in the PDF data is minimal, as evidenced by the lack of long-range ordering in both the X-ray and neutron PDFs displayed in Figures 2b and 2c, respectively. Furthermore, Figures 2b and 2c show that the CMAS glass contains obvious short- ($< \sim$3 Å) and medium-range ($\sim$3-10 Å) structural ordering, which is consistent with the structural features of silicate glasses.[10, 19, 24, 83, 84] Note that the peaks below $r \approx 1$ Å in Figures 2b and 2c are artifacts due to statistical noise, data termination errors and imperfect corrections.[67]

The nearest neighbor correlations at ~1.62, ~2.00, ~2.35 and ~2.67 Å can be assigned unambiguously to Si/Al-O, Mg-O, Ca-O and O-O correlations, respectively, based on literature data on aluminosilicate glasses.[85] However, assignment of the atomic correlations above ~3 Å for an amorphous material is difficult without an appropriate structural model (the structural representations generated in this investigation will be used to assign these atomic correlations later in the manuscript). It is noted that the X-ray and neutron data are complementary: the atom-atom correlations involving heavier elements (e.g., calcium-calcium and calcium-silicon) are more strongly weighted in the X-ray data whereas correlations involving oxygen (e.g., oxygen-oxygen, calcium-oxygen, silicon-oxygen) dominate the neutron data due to the large neutron scattering length of oxygen together with its relative abundance in the sample. Nevertheless, since the X-ray PDF data show more features compared with the neutron data, especially between 3 and 10 Å, the X-ray PDF data have been weighted more heavily when evaluating the level of agreement between the simulated data from the structural representation and the experimental PDF data.

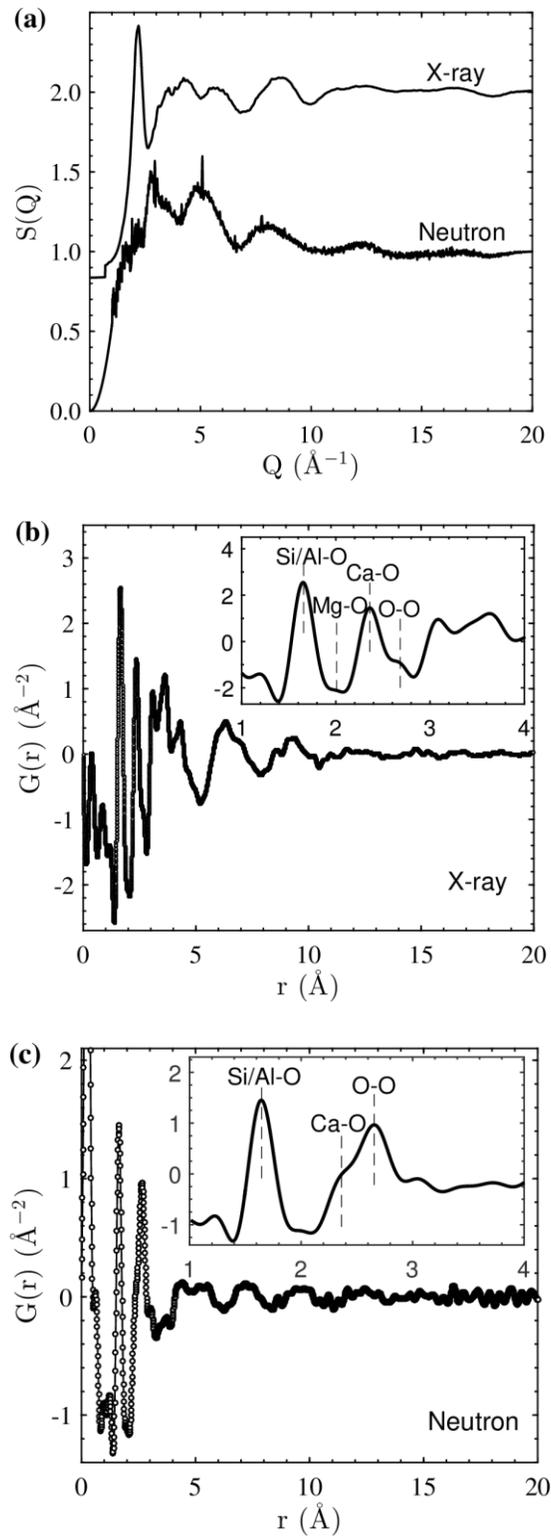

Figure 2 (a) Stacked plot of the X-ray and neutron total scattering functions, (b) X-ray PDF, and (c) neutron PDF of the CMAS glass. Inset figures in (b) and (c) show a zoom of the PDF over an *r* range of 1-4 Å.

## 3.2 Generation of the Structural Representations

### 3.2.1 Melt-quench using MD simulations

The ten amorphous structural representations for the CMAS glass were generated from five initial random structures following a commonly used melt-quench approach employing MD simulations,[38, 62, 63] as outlined in Section 2.2. Figures 3a and 3b display the level of agreement achieved between (i) the simulated PDFs of a structure (before and after the MD melt-quench process) and (ii) the corresponding experimental X-ray and neutron PDF data, respectively. As expected, the initial random structure exhibits poor agreement with the experimental PDF data ($R_w$ = ~1) and does not contain any obvious structural ordering. After being subjected to the melt-quench process using the MD simulations, the resulting structure has an improved agreement with both the X-ray and neutron experimental PDF data, where the $R_w$ values decrease to ~0.48 and ~0.31, respectively. It is seen that the MD-generated structure captures the amorphous nature of the CMAS glass, specifically by the significant decrease in intensity beyond 4 Å for both the X-ray and neutron simulated PDFs.

The local structural ordering (< ~3 Å), i.e., Si-O, Al-O, Mg-O, and Ca-O correlations, seen in the experimental data in Figure 3 is well captured by the MD-generated structure. The corresponding interatomic bond distances for the experimental data along with the MD structures (averaged over all the ten structural representations) are presented in Table 2, where it is clear that the simulation-derived distances are within 3% of the experimental values for aluminosilicate glasses.[10, 18, 19, 27, 85] The MD-derived Ca-O distance (~2.42 Å) has the largest deviation from its corresponding experimental value (~2.35 Å, see the inset in Figure 2b). Although similar Ca-O distances (2.41-2.42 Å) have been reported for MD simulations on calcium silicate glass using the same force-field,[83] the subsequent DFT calculations discussed below are seen to correct this disagreement between experiments and simulations. The double peak between 3 and 4 Å, which is attributed to the nearest-neighbor F/M-F/M correlations (network-former F = Si/Al; network-modifier M = Mg/Ca; potential correlations include Si-Si, Si-Al, Al-Al, Ca-Si, Ca-Al, Ca-Mg, Ca-Ca, Mg-Mg),[59] is also captured by the simulated PDFs to a certain degree. However, the simulated intensity of this double peak is significantly lower than the experimental data (Figure 3a). There are a number of reasons that could cause the discrepancies outlined above, including (i) the accuracy of the force-field, (ii) the finite size of the simulation cell, and (iii) the rapid cooling rate adopted in the MD simulation (~ $10^{12}$ K/s) as compared with that for typical experimental condition (1-100

K/s [63]). The impact of these factors on glass structures have been extensively investigated for different silicate glasses in the literature.[27, 31, 62, 63, 86] Furthermore, experimental contributions such as the presence of a small crystalline impurity and other trace elements (e.g., Fe, Ti and S) will also impact the level of agreement achieved between the simulated and experimental PDFs.

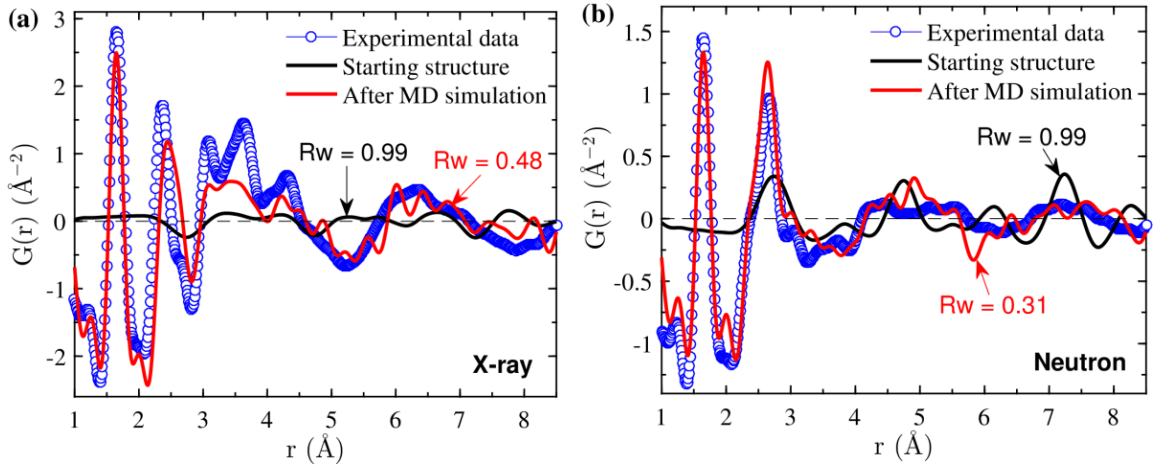

Figure 3. Simulated (a) X-ray and (b) neutron PDFs for the starting random structure (black line) and the structure after the MD melt-quench process (red line), as compared with the experimental PDF data (blue marker).

### 3.2.2 Impact of DFT geometry optimization

All ten structural representations obtained from the MD melt-quench process were subsequently subjected to DFT geometry optimization. As illustrated in Figure 4a for one structural representation, the structure shows improved agreement with the X-ray experimental PDF data after undergoing the DFT calculation. The $R_w$-value is seen to decrease from 0.48 to 0.35, and the magnitude of reduction in $R_w$ (i.e., the extent of improvement) is similar for all ten structural representations as shown in Figure S3 in the Supporting Information. Comparison of Figures 3a and 4a reveals that the lower $R_w$ value after the DFT calculation is mainly attributed to (i) an improved fit of the nearest F/M-F/M correlations between 3 and 4 Å, and (ii) a more accurate estimation of the Ca-O bond distance (i.e., 2.35 Å, as seen in Table 2). These results show that the DFT calculations lead to a better estimation of both the short- and medium-range atomic ordering in the X-ray PDF data compared with the force-field MD simulations for the CMAS glass investigated here. Nevertheless, the neutron PDF data in Figure 4b show a slight worsening of agreement ($R_w$ increases from 0.31 to 0.36) after the DFT calculation. This is mainly attributed to a slight overestimation of the O-O distance in the DFT calculation (Table 2), which has been strongly weighted in the neutron data. This agrees with previous DFT calculations on silica glass in the literature where the PBE functional has been shown to give slight overestimation of the O-O distance.[87]

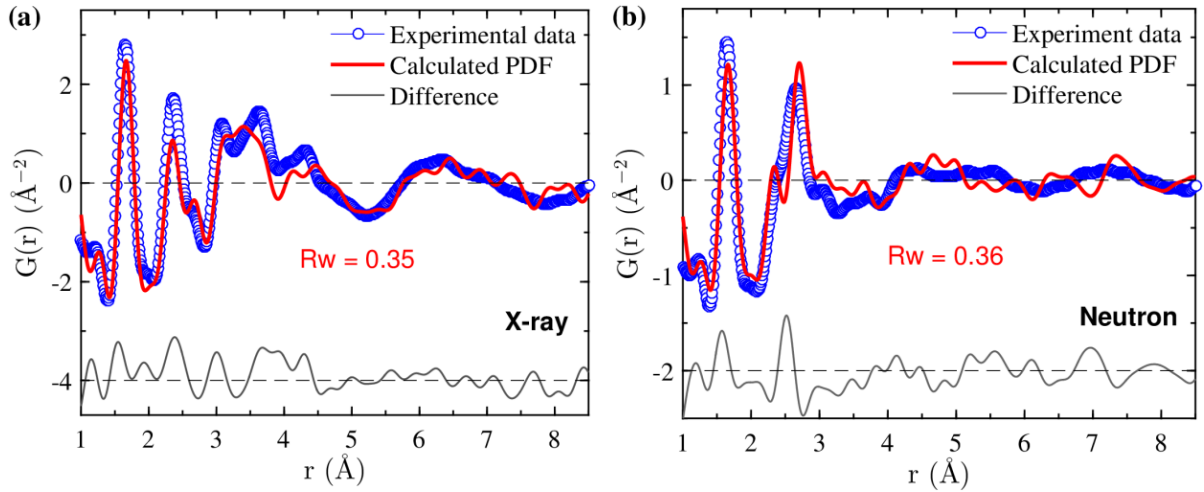

Figure 4. Calculated (a) X-ray and (b) neutron PDFs for a CMAS glass structural representation subjected to DFT geometry optimization, as compared with the experimental PDF data.

Table 2. Comparison of MD and DFT predicted interatomic distances (averaged over ten structural configurations) with the experimental values reported for different silicate glasses. The values in the brackets are one standard deviation, based on the results from the ten structural representations.

| Nearest interatomic distance (Å) | MD | DFT | Experimental PDF data in this study | Experimental data in the literature | Difference [#] (%) |
|---|---|---|---|---|---|
| Si-O | 1.63(0.00) | 1.64(0.00) | 1.64[†] | ~1.61-1.64 [10, 18, 19, 27, 85] | 0 |
| Al-O | 1.75(0.01) | 1.77(0.01) | N/A | ~1.74-1.77 [10, 18, 19, 27, 85] | 0.9 |
| Mg-O | 2.03(0.02) | 2.02(0.02) | 2.00[*] | ~2.00 [85] | 1.0 |
| Ca-O | 2.42(0.01) | 2.35(0.01) | 2.35[*] | ~2.34-2.36 [10, 18, 19] | 0 |
| O-O | 2.68(0.01) | 2.71(0.01) | 2.66[†] | ~2.65-2.67 [19, 85] | 1.9 |

[†] Derived from neutron PDF data

### 3.2.3   Iterative RMC-DFT process

To assess if the structural representations obtained in Section 3.2.2 could be further improved, specifically by allowing them to further explore the potential energy surface of the CMAS glass, one structure was subjected to the iterative RMC-DFT process outlined in Section 2.2, with the results given in Figure 5. The level of agreement between the simulated X-ray PDF of the structure and the experimental X-ray PDF data throughout the iterative process is shown in Figure 5a, where the agreement with the experimental data improves after RMC refinement of all the atomic positions (comparing "1st DFT" and "1st RMC" in the figure). However, as expected, the structural representation at this point is chemically implausible, containing non-physical molecular arrangements such as distorted Si tetrahedra and Mg polyhedra (refer to Figure S4 and Table S1 in the Supporting Information for the details). A single point energy calculation of the RMC-refined structure using DFT could not reach convergence since the RMC-refined structure is far away from a local minimum on the potential energy surface. The RMC refinement process is known to give the most disordered structural representations that are in agreement with the experimental data.[82] Constraining the data during the refinement and using a large simulation box are suggested methods to limit the chemical implausibility generated by RMC.[82]   In this investigation we have employed the iterative process, and specifically DFT calculations, to regain chemical plausibility, where after another round of the DFT calculation ("2nd DFT" in Figure 5a), the structure no longer contains distorted Si tetrahedra and other non-physical arrangements and converges at a ground-state energy of ~$-3230$ eV (Figure 5b). The partial atom-atom correlations throughout the RMC-DFT process together with changes in bond lengths and coordination numbers are given in Figure S4 and Table S1 of the Supporting Information.

Throughout the iterative process the level of agreement ($R_w$) between the experimental and simulated X-ray PDFs oscillates between ~0.2 and ~0.35 (after the RMC simulation and the DFT calculation, respectively), where the $R_w$ value of 0.35 for the DFT calculation is reached prior to any RMC simulations. In fact, Figure 5b shows that several rounds of RMC and DFT calculations do not alter the ground-state energy of the DFT-derived structure (to within 0.5 eV or 0.015% of the total energy). There are slight changes to the local atomic structure of the CMAS glass during the iterative process, specifically to the calcium ions where a slight reduction of the average coordination number occurs (from 6.80 to 6.77, see Table S1 of the Supporting Information). However, the small change in coordination number is likely attributed to the relatively small box size used here (due to computational requirements of the DFT calculations), since, as discussed in the next section, analysis of the ten structural configurations obtained using the MD melt-quench process followed by DFT reports an average calcium coordination number of $6.73 \pm 0.07$. Therefore, the structures obtained after a single round of DFT geometry optimization are considered as the final structural representation for the CMAS glass investigated here and will be analyzed in detail in the next section. This finding is in contrast to previous investigations[73, 74] using the iterative process, where several rounds of experimental refinement followed by DFT calculations or MD simulations were required to obtain a structure that was experimentally plausible (as assessed using $R_w$) and chemically feasible (at a minimum on the potential energy surface). Hence, the starting structures generated in this study using the amorphous prebuilder followed by the melt-quench force-field MD process are in the vicinity of the experimental local minimum for this CMAS glass, and therefore experimental refinement of the structural representations is not needed.

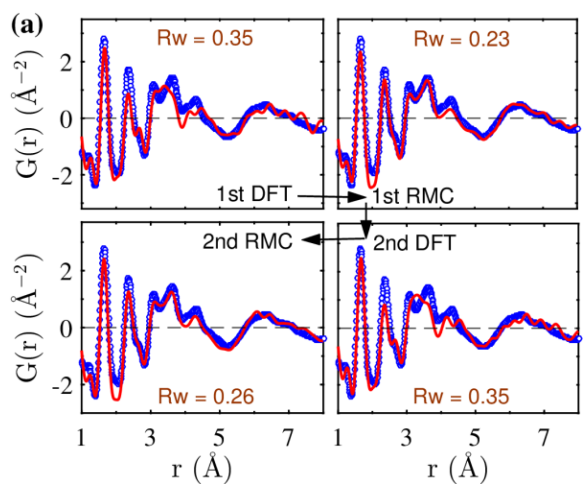

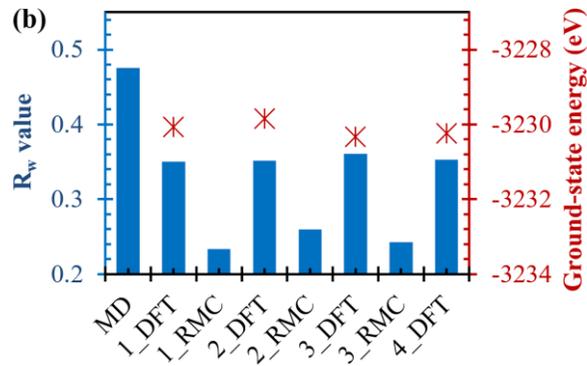

Figure 5. (a) Comparison of the experimental X-ray PDF of the CMAS glass with the simulated X-ray PDF of the structural representation throughout the iterative RMC-DFT process (i.e., after DFT geometry optimization or RMC refinement). (b) Evolution of the $R_w$ value (agreement between simulated and experimental X-ray PDF) and the ground-state energy of the structural representation throughout the iterative RMC-DFT process.

### 3.3    Analysis of the Final Structural Representations

Figure 6a displays a typical final structural representation for the CMAS slag obtained after a single round of DFT geometry optimization, which clearly shows the amorphous nature of the structure. In general, the structure can be described largely as a depolymerized chain-like network structure consisting of corner-sharing $SiO_4$ and $AlO_4$ tetrahedra.[88] The aluminosilicate network in Figure 6b reveals a considerable amount of Al-O-Al linkages, which will be quantified along with other structural features in the following subsections. Note that all the structural features and properties reported below are based on analysis of ten final structural representations (each obtained after a single round of DFT geometry optimization).

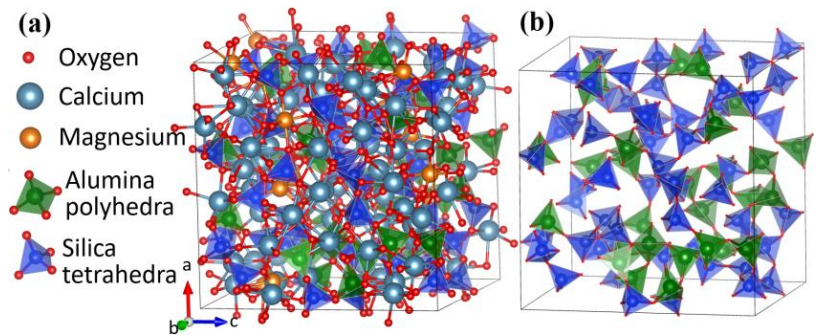

Figure 6. (a) A representative final structure of the CMAS glass obtained after a single round of DFT geometry optimization. (b) The aluminosilicate network of the CMAS glass structure in (a).

### 3.3.1 Coordination states

The evolution of coordination number (CN) with increasing cutoff distance for the different atom-atom pairs is illustrated in Figures 7a and 7b. It is clear from Figure 7a that the first coordination shells (involving oxygen atoms) of Si and Al are well defined since a plateau is reached for the CNs by 1.8 and 1.9 Å, respectively. The Si atoms are 100% tetrahedrally coordinated (see Table 3) while Al atoms are seen to be dominated by tetrahedral coordination with a small proportion of V-fold coordination (~3 %), as shown in Table 3 (see Figure S5a of Supporting Information for CN distribution of Si and Al). For the CMAS glass investigated here, there are excess Ca and Mg atoms in the structure beyond those required to charge-balance the negative tetrahedral alumina sites, and therefore there should not be any V-fold Al atoms in the system based on simple stoichiometric considerations.[22] A previous NMR ($^{27}$Al) study on a CMAS glass with a similar chemical composition also suggested a single IV-fold coordination state for all Al atoms.[88] However, there are many MD simulations and experimental data (including $^{27}$Al NMR) on peralkali/peralkaline-earth aluminosilicate glasses (e.g., $Na_2O$-$Al_2O_3$-$SiO_2$,[89] CAS,[10-12, 15, 28, 34, 35] $MgO$-$Al_2O_3$-$SiO_2$ (MAS),[34, 35, 85, 90, 91] and CMAS[34-36]), where a small proportion of Al species with higher coordination states have been identified, in contrast to what is expected from consideration of the stoichiometry. In addition, it has been shown that cations with high field-strength (e.g., $Ca^{2+}$, $Mg^{2+}$) often lead to an increase in the amount of Al that has a CN above four (as compared to low field-strength cations, e.g., $Na^+$, $K^+$),[35, 36] hence it is possible for V-fold Al to form in the CMAS glass studied here.

In contrast with the evident cutoff distances for the average CNs of Si and Al atoms seen in Figure 7a, the CNs for Mg and Ca atoms (with oxygen) are highly dependent on the selected cutoff distance, which might contribute to the different oxygen CNs reported in the literature for Mg (~4-7[85, 92-94]) and Ca (~5-9[10, 16, 34]) atoms in silicate glasses from simulations in comparison with experimental data. By using cutoff distances corresponding to the first minimum after the main peak of the partial PDFs, we see in Table 3 that the average CNs of Ca-O and Mg-O in the current work are approximately 6.73 and 5.15, respectively, which are in agreement with the previously reported values obtained using simulations and experiments (Table 3). Even at these fixed cutoff distances (Table 3), both Mg and Ca atoms have a distribution of oxygen CNs, as illustrated in Figure S5 in the Supporting Information. Figure S5b shows that the Mg environment in the CMAS glass consists of 4-, 5-, 6-, and 7-fold coordinated sites, with 5-fold dominating as confirmed using

XANES,[85, 92] X-ray/neutron diffraction coupled with RMC refinement,[85] and MD simulations.[58] Nevertheless, previous NMR ($^{25}$Mg) studies on MAS and CMAS glasses shows Mg is mainly in 6-coordination.[88, 94] The discrepancy between different experimental results is partially attributed to the sensitivity of different experimental techniques to specific Mg bonding environments, as has been discussed in ref. [95] for XANES and NMR.

The local coordination environment of the Ca atoms in the CMAS glass is dominated by 6- and 7-fold coordinated Ca, along with the presence of 5-, 8-, and 9-fold coordination states (see Figure S5b in the Supporting Information for details). These results agree with previous experimental and MD studies where Ca has been shown to mainly reside in distorted sites with six to seven oxygen neighbors.[10, 16, 34, 59, 88] It is noted that literature data on alkaline-earth silicate glasses generally conclude that Ca atoms have higher CNs than Mg atoms within their first coordination shell, which is mainly attributed to the lower field-strength (defined as $Z/d^2$, where $Z$ is the cation charge and $d$ is the cation-oxygen distance) of Ca cation (~0.36) as compared to Mg cation (~0.46-0.53).[94]

To evaluate whether there is a preference for a specific network-modifier (i.e., Ca and Mg) to charge-balance Al polyhedra, we have calculated the average number of Ca and Mg atoms around Si and Al atoms as a function of cutoff distance, as shown in Figure 7b. It is clear that the Ca (or Mg) CNs around Si and Al atoms are similar, indicating no obvious preference for Ca (or Mg) to associate with Si or Al atoms. The evolution of the (Ca CN)/(Mg CN) ratio around Si or Al as a function of the cutoff distance (Figure 7c) shows that this ratio is slightly higher than the overall Ca/Mg compositional ratio (~5.9) of the CMAS glass at a cutoff distance of 4-5 Å, where the first coordination shells between the network-formers and network-modifiers are located. At the fixed cutoff distances for each of the Ca/Mg-Si/Al pairs (as shown in Table 3), we get a (Ca CN)/(Mg CN) ratio of ~7.4 and ~7.5 around Si and Al atoms, respectively. At a cutoff distance larger than 5-6 Å, the (Ca CN)/(Mg CN) ratio approaches the overall Ca/Mg ratio of the sample. These results indicate that there is a slight preference for Ca cation (over Mg) to associate with both types of network-formers (i.e., Si and Al) within their first coordination shells, however, the Ca-Mg mixing around Al and Si atoms becomes completely random outside the first coordination shells. The same features are also seen around 5-fold Al site (as shown in Figure 7c), which also suggest its slight preferential proximity with Ca (over Mg) atom. The cause of this slight preferential proximity of Ca with Si, Al and 5-fold Al will be touched on in Section 3.3.3. It is noted that this observation is different from a previous MD investigation,[59] which showed that the (Ca CN)/(Mg CN) ratios around Si (1.6) and Al (3.7) are significantly lower than the overall Ca/Mg compositional ratio of the CMAS glass in that study (4.4), indicating a preferential association of Mg with both Si and Al.

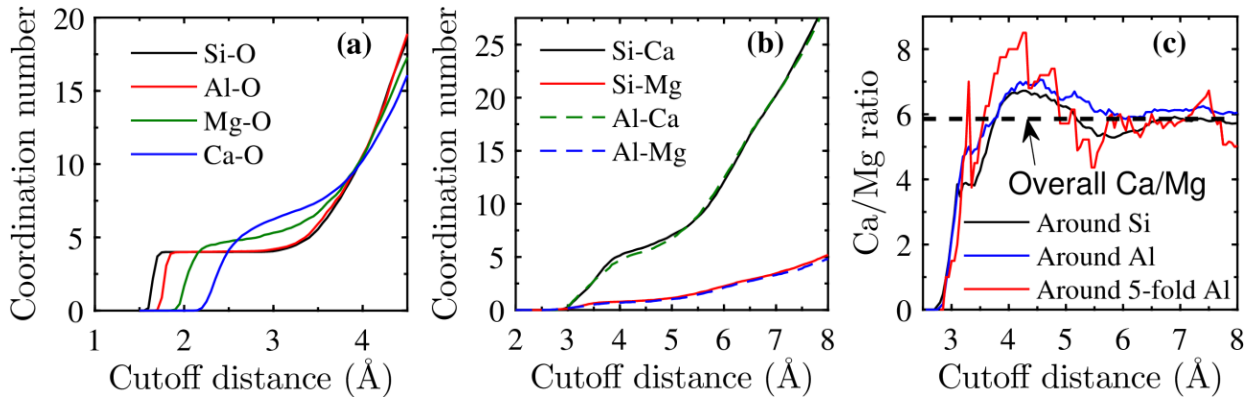

Figure 7. Evolution of coordination number as a function of cutoff distance for (a) Si/Al/Mg/Ca-O (i.e., number of oxygen atoms surrounding Si, Al, Mg, Ca), (b) Si-Ca, Si-Mg, Al-Ca and Al-Mg (i.e., number of Ca, Mg atoms surrounding Si, Al), and (c) Ca/Mg CN ratio around Si, Al, and 5-fold Al atoms. The y-axis Ca/Mg ratio in (c) is calculated using the data in (b), for example, the Ca/Mg ratio around Si is determined by (Ca CN around Si)/(Mg CN around Si) at each given Ca/Mg-Si cutoff distance. The results are averages based on the ten structural representations optimized using DFT calculations.

Table 3. Coordination numbers at fixed cutoff distances for different atom-atom pairs. For an X-Y atom-atom pair, the coordination number of X, averaged over the ten DFT-optimized structural representations, is given, along with one standard variation shown in the brackets. Literature data on different types of silicate glasses (e.g.,CAS[6, 16], MAS,[85, 92] MS,[94] CAMS,[88] and NCAS[54]) are also given for comparison.

| Atom pairs | Cutoff distance (Å) | Current study | Literature data | |
|---|---|---|---|---|
| | | | Experiments (e.g., NMR[88, 94], XANES[16, 54, 92], and neutron diffraction[6, 10]) | Simulations (e.g., MD[6, 59] and RMC+X-ray/neutron diffraction[85, 93]) |
| Si-O | 2.2 | 4.00 (0.00) | 4,[88] 4.04[10] | 4[6, 59, 85] |
| Al-O | 2.5 | 4.03 (0.03) | 4,[88] 4.1-4.20[6, 10] | 4.0-4.10,[6, 59] 4.1-4.16[85] |
| Mg-O | 2.9 | 5.15 (0.15) | 5,[92] 6,[88, 94] | 4.5,[93] 4.75-5.09,[85] 5.5,[59] 4.75-5.09[85] |
| Ca-O | 3.2 | 6.73 (0.07) | 7,[54, 88] 6-7[16] | 6.00-6.24,[6] 6.7,[59] 7-7.5[54] |
| Si-Ca | 4.5 | 5.96 (0.10) | N/A | 5.6[59] |
| Si-Mg | 4.1 | 0.81 (0.03) | N/A | 3.5[59] |
| Al-Ca | 4.5 | 5.58 (0.28) | N/A | 5.2[59] |
| Al-Mg | 4.2 | 0.74 (0.20) | N/A | 1.4[59] |

### 3.3.2 Oxygen environment

The oxygen environment, and in particular, the proportion of NBO species, has a large impact on glass properties (e.g., hardness,[9] chemical reactivity,[44] durability,[8] and glass transition temperature[7]). Hence, we have calculated the proportion of different types of oxygen species based on the ten DFT-optimized structural representations and the results are shown in Figure 8. It is seen that the CMAS glass studied here has an NBO content of ~ 58.9% (percent relative to total amount of oxygen atoms), which gives an NBO/T (T = Si or Al tetrahedra) of 1.75, a reflection of the degree of depolymerization of the glass structure. This indicates that this CMAS glass has, on average, a close to short-chain structure, which is consistent with NMR measurements on a CMAS glass of similar composition (with an average $Q^n$ species of n =2.2).[88] Nevertheless, as shown in Table 4, this percentage is slightly lower than the theoretical NBO content (~64.6%) estimated using simple stoichiometry arguments[27] and assuming that the glass system consisted of perfect tetrahedra with only two-fold oxygen atoms (i.e., no free oxygen (FO) that are not connected with any network formers or tri-cluster oxygen (TO) connected with three network formers). Similar underestimation of the NBO content has been reported in a MD study for CMAS melts with similar compositional ranges as the current study,[58] nevertheless, studies on CAS glasses[21, 28, 29] have often exhibited higher NBO contents than the theoretical estimation. The discrepancy is mainly attributed to the fact that a small proportion of FO and TO are regularly observed in aluminosilicate glasses[27, 28, 58, 89], as also shown in the current study (Figure 8), and their proportion varies considerably depending on the glass composition. The underestimation of the NBO percent in the current study is partially due to the relatively high proportion of FO (~2.5 %, as compared to TO of ~0.2 %), arising from the relatively high modifier content (at ~50 %, amount of Ca and Mg relative to Ca, Mg, Si and Al). In contrast, the CAS glasses in refs. [21, 28, 29] have much higher proportions of TO (~3-7%) due to their relatively lower modifier content (~10-30 %) and/or higher

Al/Si ratio (>>1), which may have led to the higher observed NBO content mentioned above (as compared to estimation from simple stoichiometric argument).

The local environment surrounding the NBO sites has also been analyzed and reported in Figure 8 and Table 4, where it is seen that the proportion of NBO associated with Si atoms is about 4 times higher than that associated with Al atoms, in contrast with the overall Si/Al compositional ratio in the CMAS glass (i.e., 2.1 Si atoms for every Al atom). This suggests that there is a preferential formation of NBO around Si atoms and BO around Al atoms, which is consistent with previous studies on aluminosilicate glasses,[14, 16, 17, 28, 36, 85] where Al atoms are shown to prefer to reside in more polymerized environments than Si atoms. DFT calculations have shown that these preferential associations are mainly attributed to the higher energy penalty for the formation of Al-NBO (108 kJ/mol) as compared to Si-NBO (72 kJ/mol).[13]

With respect to the BO sites, Figure 8 shows that there is a small proportion of Al-BO-Al linkages, indicating that the Al-O-Al avoidance principle (Loewenstein's rule) prevalent in crystals is not fulfilled in this CMAS glass, as has been previously reported in numerous studies on aluminosilicate glasses, including simulations[27, 28] and experiments.[20, 36] Interestingly, the proportion of Al-BO-Al linkages (5.5 %) is seen to be even higher than that estimated from a complete random distribution of Si and Al atoms around BO sites (3.6%, as shown in Table 4; detailed calculations are shown in the Supporting Information). This result indicates that the Al-O-Al avoidance principle is violated for this CMAS glass, which is different from several previous studies on NAS and CAS glasses, where the Al-O-Al avoidance principle is only partially violated.[20, 27] The difference may be attributed to the higher proportions of strong modifier cations ($Ca^{2+}$, $Mg^{2+}$) in the CMAS glass studied here (as compared to the NAS and CAS glasses in refs. [20, 27]), since the high field strength cations favor the negative charge concentration (e.g., Al-BO-Al) more than low-strength modifier cations (e.g., $Na^+$) and hence promote the formation of Al-BO-Al linkages.[20, 21, 91] This is supported by another MD study on CAS glasses, which showed that the proportion of Al-BO-Al sites becomes higher than theoretical values (assuming random distribution of Si and Al atoms around BO) when the Ca content reaches 50%.[28]

Finally, Table 4 shows that there is a preferential intermixing of Si-Al around BO, as evidenced by the higher proportion of Si-BO-Al (~20.7 %) and lower proportion of Si-BO-Si (~12.3 %) linkages than the theoretical estimation based on random mixing of Si and Al (~15.9 and ~16.4 %).

This preferential Si-Al intermixing is consistent with previous studies on aluminosilicate glasses where mixing between different network-formers (as opposed to the same type of network-formers) are preferred.[36] This is likely attributed to the often observed negative enthalpy of mixing between Al-rich and Al-poor glass, as has been shown in a solution calorimetry study on CMAS glasses.[23, 96] Hence, the deviation from the theoretical proportions of oxygen species (Table 4) are driven by two competing mechanisms: (i) strong modifier cations (i.e., $Ca^{2+}$, $Mg^{2+}$) promote the formation of more negative BO sites (Al-O-Al > Si-O-Al > Si-O-Si) and (ii) negative enthalpy of mixing promote intermixing of Si and Al atoms (Si-O-Al > Si-O-Si, Al-O-Al). The high proportion of Ca+Mg atoms in the CMAS glass (over 50%) renders mechanism (i) as the dominant mechanism, resulting in the larger proportion of Al-O-Al linkages than theoretically estimated. Both mechanisms (i) and (ii) favor formation of Si-O-Al over Si-O-Si, leading to the higher proportion of Si-O-Al and lower proportion of Si-O-Si than theoretical estimations (Table 4).

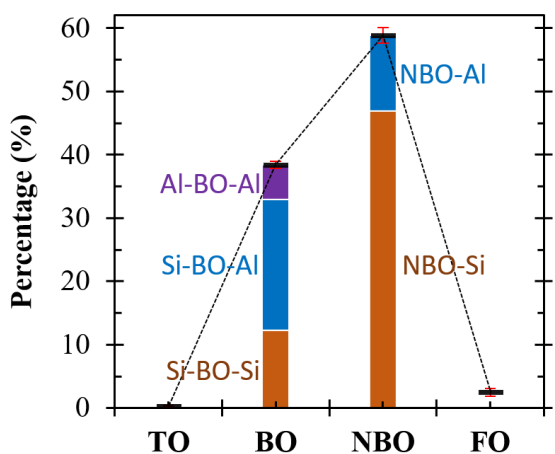

Figure 8. Proportions of the different types of oxygen species. The total percentages of tri-cluster oxygen (TO), bridging oxygen (BO), non-bridging oxygen (NBO) and free oxygen (FO) are averages based on the ten DFT-optimized structural representations, with the red error bar indicating one standard deviation.

Table 4. Comparison of the proportions of different types of oxygen species (bridging oxygen (BO) and non-bridging oxygen (NBO)) between the structural representation in the current study and the theoretical estimation based on simple stoichiometric considerations and random mixing of network-formers and oxygens (refer to the Supporting Information for the calculations). The average values obtained for the ten structural representations are given along with one standard deviation.

| | Percentage of different types of oxygen species (%) | | | | | | |
|---|---|---|---|---|---|---|---|
| | BO | NBO | NBO-Si | NBO-Al | Si-BO-Si | Si-BO-Al | Al-BO-Al |
| Theoretical estimation | 35.9 | 64.1 | 43.4 | 20.6 | 16.4 | 15.9 | 3.6 |
| Structural representations | 38.4(0.5) | 58.9(1.3) | 46.9(2.1) | 12.0(2.0) | 12.3(1.1) | 20.7(1.8) | 5.5(1.4) |

### 3.3.3   Distribution of modifier cations around different oxygen species

Figure 9 shows the number of modifier cations (i.e., Ca and Mg) around the different types of oxygen species within their first coordination shell (based on analysis of the ten DFT-optimized structural representations), where the local Ca/Mg ratio around each type of oxygen species is compared with the average Ca/Mg ratio around O atoms (i.e., Ca/Mg of 7.6) and the overall Ca/Mg compositional ratio in the CMAS glass (i.e., Ca/Mg of 5.9). The difference between the two Ca/Mg ratios is attributed to the difference in the average oxygen CNs of Ca and Mg as seen in Table 3 (7.6/5.9 = ~1.3 = 6.73/5.15). In general, the average number of modifier cations (both Ca and Mg) increases as the number of network formers around the oxygen site decreases (i.e., number of modifier cations increases as transition from TO to BO, NBO and FO sites). This is expected since more cations are required for charge-balancing as the oxygen sites become increasingly negative. Previous [17]O NMR measurements on CMAS glasses suggested a prevalence of 3Ca-NBO-Si around NBO-Si sites.[23] This observation is generally consistent with our results in Figure 9, where an average of ~2.6 Ca atoms are seen around the NBO-Si sites with 3Ca-NBO-Si as the dominant species (see Figure S6 in the Supporting Information).

Furthermore, the Ca/Mg ratio around the oxygen site is seen to decrease as the oxygen site becomes increasingly negative (Figure 9), which is attributed to the higher field-strength of Mg (as compared to Ca), rendering it more effective in charge-balancing the more negative oxygen site. It is also seen that the Ca/Mg ratios around the three BO sites (~9.0-15.5) are higher than the average Ca/Mg ratio around all O atoms (~7.6), while the Ca/Mg ratios around the two NBO sites (~7.0) are slightly lower than this average value. This result reveals a slight preference for Ca atoms to compensate charge and for Mg atoms to modify aluminosilicate network (creating NBO) in the CMAS glass, which is consistent with the known preference of a high-field strength cation to associate with NBO (e.g., preferential association of Ca with NBO for Ca-Na[54, 97], and Mg with NBO for Mg-K[98]).

The lowest Ca/Mg ratio (~4.6) is seen around the FO sites (Figure 9), which are the most negative oxygen sites in the system, indicating a strong preferential association of FO sites with the Mg atom (as opposed to Ca). Again, this preference is attributed to the higher field strength Mg atom which enables it to more effectively charge-balance the highly concentrated negative charge surrounding FO sites. It has been previously shown that the FO content in CAS glasses with network-modifier molar contents of ~55-61% is around ~0.5-1.0%[28], which is much lower than the FO content in the current study (i.e., ~2.3%), although the network-modifier content in the CMAS glass studied here is lower (~50%). Moreover, a recent study on CAS and MAS melts (at 1773 K) showed that the MAS melt has a much higher FO content than the corresponding CAS melt for the same amount of modifiers (i.e., Mg or Ca).[55] These results suggest that the presence of Mg in CMAS glass promotes the formation of FO, which is a contributing factor to the underestimation of the NBO content in the simulation as compared to the simple stoichiometric calculation (as shown in Table 4). This preferential association of Mg atoms with FO sites also explains the higher Ca/Mg ratios around Si and Al atoms within their first coordination shell as compared to the average Ca/Mg compositional ratio in the CMAS glass (Figure 7 and Table 3). Furthermore, since FO sites are the most reactive oxygen sites, they are more prone to dissolve in aqueous solutions, which could be a major reason why CMAS glasses with higher Mg contents have been shown to exhibit higher reactivity.[39, 42] Nevertheless, a carefully designed study is warranted to further confirm the positive correlation between Mg and FO content for CMAS glasses at room temperature. It is also noted that CMAS glass reactivity in an alkaline solution is highly complex and other factors, such as NBO content, particle size distribution and thermal history of the CMAS glass can also have a large impact on its reactivity.[44]

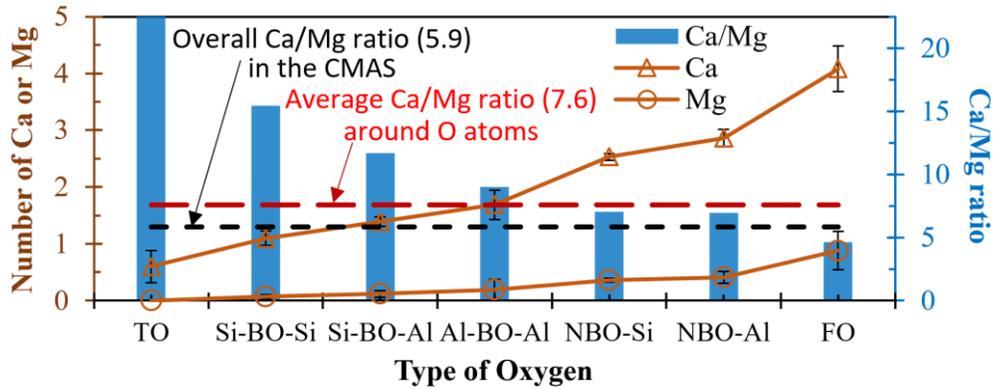

Figure 9. Average number of Ca or Mg around each type of oxygen species, where TO, BO, NBO and FO denote tri-cluster oxygen, bridging oxygen, non-bridging oxygen and free oxygen, respectively. The cut-off distances for Ca-O and Mg-O are fixed at 3.2 and 2.9 Å, respectively.

Finally, the deviation of the Ca/Mg ratios around the different oxygen sites from the average ratio indicates a non-random distribution of Ca-Mg around the oxygen sites with a slight degree of segregation (i.e., separate clustering of Ca and Mg atoms), which has been suggested for Ca-Mg around NBO sites in CMAS glasses according to an $^{17}O$ NMR study.[23] The clustering of Mg atoms is clearly evident in Figure 10a, where a typical CMAS structural representation exhibits formation of small Mg clusters, with Mg-Mg pairs with distance smaller than 3.5 Å highlighted using red dashed circles. This is further supported by the Mg-Mg partial correlation averaged over the ten final structural representation (Figure 10b), which exhibits two peaks located at ~2.8 and 3.3 Å. These distances are much smaller than the theoretical distance of ~7.5 Å assuming a random distribution of Mg atoms in the unit cell, which confirms clustering of Mg atoms in the CMAS glass to a certain extent.

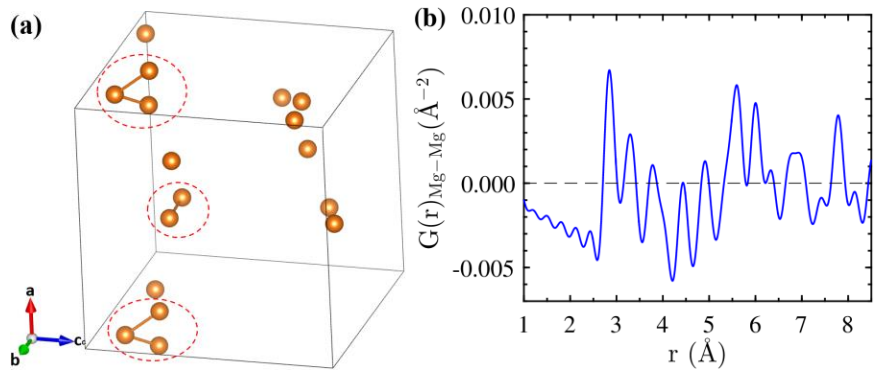

Figure 10. (a) Small clusters of Mg atoms in a typical CMAS structural representation, and (b) Mg-Mg partial X-ray PDF calculated using the ten final structural representations. For clarity, only Mg atoms are shown in (a) and Mg-Mg pairs with distance smaller than 3.5 Å are highlighted using red dashed circles.

### 3.3.4   Partial PDFs

The results presented in the previous sections show that the structural representations generated for the CMAS glass using the MD melt-quench process followed by DFT geometry optimization not only agree with our X-ray and neutron scattering data but also are generally consistent with literature data on aluminosilicate glasses, specifically in terms of interatomic distances, coordination numbers, and oxygen environments. With these realistic structural representations, it is now possible to unambiguously assign the features seen in the experimental PDF data (Figures 2b and 2c), which would otherwise be an extremely challenging task to perform for the medium-range ordering (~3-8 Å) due to the overlapping nature of many individual atom-atom partials.

Figure 11 shows the partial X-ray PDFs based on the ten structural representations that have been subjected to one round of DFT geometry optimization. It is seen that the medium-range ordering between ~4-5 Å is mainly attributed to the second nearest Si-O and Ca-O correlations in the CMAS glass, whereas the medium-range ordering between ~5-8 Å is mainly due to the third nearest Ca-O correlation and the second nearest Ca-Ca and Ca-Si correlations. Previously, the X-ray PDF peak located at ~3 Å for CMAS glasses has been assigned primarily to the nearest Si-Si/Al correlations based on partial radial distribution functions,[37, 59] however, Figure 11 shows that this peak is dominated by the nearest Ca-Si/Al correlations with only minor contributions from the nearest Si/Al-Si/Al correlations. Another mis-assignment in ref.[37] is the shoulder at ~3.3 Å (as seen in the inset figure in Figure 2b), which was assigned to the nearest Mg-Si/Al correlations. However, Figure 11 clearly shows that this shoulder is mainly attributed to the nearest Ca-Ca/Si correlations, with negligible contribution from Mg-Si/Al correlations.

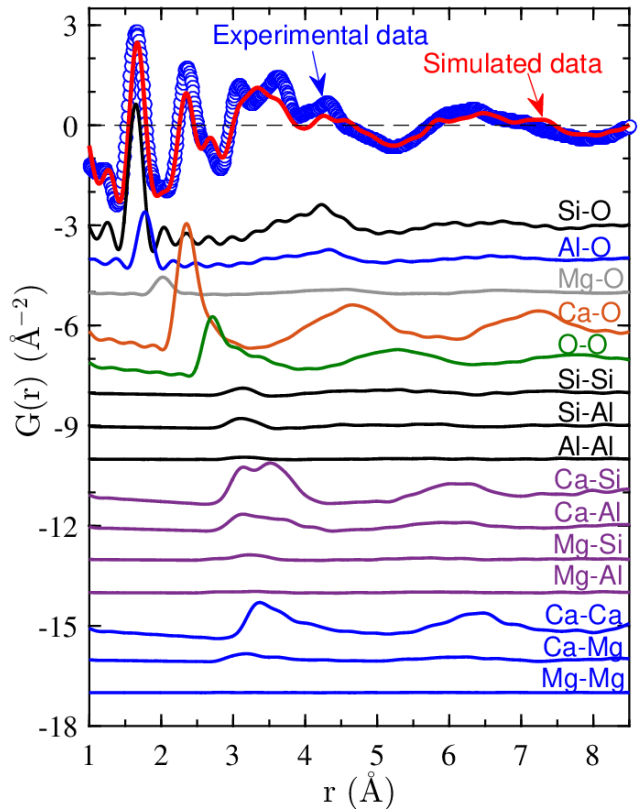

Figure 11. Simulated partial X-ray PDFs based on the ten structural representation of the CMAS glass that have been geometry-optimized using DFT calculations.

In addition to the peak at ~3.1 Å, the nearest Ca-Si/Al correlations exhibit a second peak at ~3.6 Å. This double peak feature for the nearest Ca-Si/Al correlations is commonly observed in CAS glasses,[59] and is attributed to the connectivity between Si/Al tetrahedra and Ca polyhedra, where edge-sharing connectivity leads to the peak at ~2.8-3.1 Å and corner-sharing is responsible for the peak at ~3.3-3.6 Å. This is illustrated in Figure S7 in the Supporting Information, where it is clearly seen that the corner-sharing Ca-Si/Al distances are ~0.4-0.6 Å larger than that of the edge-sharing Ca-Si/Al distances. The O-O partial PDF shows that the nearest O-O correlation exhibit two shoulders at ~3.0 and ~3.4 Å, in addition to the main peak at ~2.7 Å. Calculation of the O-O distances in all the Si/Al tetrahedra and Mg/Ca polyhedra (see Figure S8 in the Supporting Information) reveals that the main peak at ~2.7 Å is mainly attributed to the O-O distances in Si tetrahedra whereas the shoulder at ~3.0 Å is dominated by O-O distances from Al tetrahedra and Ca polyhedra. The shoulder of the O-O partial PDF at ~3.4 Å is primarily attributed to O-O correlations in the Ca polyhedra.

## 3.4 Broader Implications

In this paper, realistic structural representations for a CMAS glass system have been obtained by combining force-field MD simulations and DFT calculations with RMC refinement against X-ray and neutron scattering data. Hence, this methodology will be of interest to the amorphous materials community, especially since the establishment of composition-structure-performance relationships is an area of great interest to materials science. Although the methodology concentrates on the first part of this relationship (composition-structure), this information is paramount for subsequent structure-property investigations. Furthermore, since certain classes of materials lack accurate force-fields for MD simulations, this method will help circumvent this limitation since it combines the benefits of MD simulations (computationally efficient) and DFT calculations (more accurate). Therefore, even though the selected force-field may not be accurate enough to generate structural representations on its own for a given material, when followed with DFT calculations the resulting structures should be closer to reality, which can be verified via comparison or refinement against experimental data.

With the generation of realistic structural representations, it becomes possible to unambiguously assign the medium-range ordering generally seen in the experimental PDF data of CMAS glasses and related amorphous materials. This information is particularly useful when combined with *in situ* PDF analysis to study amorphous-amorphous transformations, such as CMAS glass dissolution in aqueous environments, where changes to individual PDF peaks during the dissolution process can be directly related to the disappearance of certain structural features in the CMAS glass. The combination of glass structure modeling with *in situ* PDF analysis will be extremely helpful for studying the kinetics and mechanisms of glass dissolution, which is crucial for a range of industrially-important processes, including bioglass dissolution, low-$CO_2$ cements formation and degradation, glass corrosion, and nuclear waste encapsulation.[44, 53, 99, 100]

## 4 Conclusions

In this study, we combined force-field MD simulations and quantum chemical calculations with X-ray and neutron total scattering experiments to generate ten realistic structural representations for a quaternary $CaO-MgO-Al_2O_3-SiO_2$ (CMAS) glass. Analysis of the data showed that a single round of DFT calculations (geometry optimizations) on the MD-generated structures is sufficient for generating structural representations with excellent agreement with both X-ray and neutron experimental pair distribution function (PDF) data, without the need for additional refinement against experimental data.

Quantitative analysis of the ten structural representations showed that the CMAS glass structures generated using the method outlined in this study not only agree with our X-ray and neutron scattering data but also are generally consistent with literature data on aluminosilicates with respect to interatomic distances, coordination numbers, and oxygen environments. Specifically, for the nearest-neighbor bonding environment with oxygen atoms Al is mainly in IV-coordination with a small proportion of V-fold, whereas Ca and Mg cations exhibit a much wider distribution of coordination states, with an average of ~6.73 and ~5.15, respectively. Analysis of the next nearest neighbors revealed that there is slight preference for Ca atoms (over Mg) to associate with both network-formers (i.e., Si and Al atoms). Analysis of the oxygen environment revealed several key features that are consistent with the literature, including violation of the Al-O-Al avoidance principle, preferential association of NBO with Si atoms (as opposed to Al atoms), and Si-Al intermixing. Calculation of the modifier environment around the different oxygen species showed a slight preference for Ca atoms to act as charge compensators and Mg atoms as network modifiers. The results also revealed a preferential association of Mg with FO sites and a tendency for Mg to from small clusters in the CMAS glass. This may help explain the higher reactivity of CMAS glass with higher Mg content when exposed to alkaline aqueous environments that has been observed in the literature.

Finally, this investigation has enabled atom-atom correlations responsible for the medium-range ordering (~3-8 Å) seen in the experimental PDF data of CMAS glass to be accurately assigned. Correct assignment of these correlations in this region will not only enable for better interpretation of existing PDF data but will also lead to advances in our understanding of dissolution mechanisms of CMAS glass (and related amorphous materials systems) in aqueous environments via experimental methods such as *in situ* PDF analysis.

## 5    Acknowledgments

This material is based on work supported by the National Science Foundation under Grant No. 1362039. The DFT calculations were performed on computational resources supported by the Princeton Institute for Computational Science and Engineering (PICSciE) and the Office of Information Technology's High Performance Computing Center and Visualization Laboratory at Princeton University. The authors would like to acknowledge the support from the technical staff associated with the research computing facility at Princeton University. KG was partially supported by a Charlotte Elizabeth Proctor Fellowship from the Princeton Graduate School. The 11-ID-B beam line is located at the Advanced Photon Source, an Office of Science User Facility operated for the U.S. DOE Office of Science by Argonne National Laboratory, under U.S. DOE Contract No. DE-AC02-06CH11357. The NPDF instrument is located at Los Alamos Neutron Science Center, previously funded by DOE Office of Basic Energy Sciences. Los Alamos National Laboratory is operated by Los Alamos National Security LLC under DOE Contract DE-AC52-06NA25396. The upgrade of NPDF was funded by the NSF through grant DMR 00-76488.

## 6    References

1.      E. Le Bourhis, *Glass: mechanics and technology*, John Wiley & Sons, Weinheim, Germany, 2 edn., 2014.
2.      B. O. Mysen and P. Richet, *Silicate glasses and melts*, Elsevier, Amsterdam, Netherlands, 2 edn., 2018.
3.      M. E. Lines, J. B. MacChesney, K. B. Lyons, A. J. Bruce, A. E. Miller and K. Nassau, *J. Non-Cryst. Solids*, 1989, **107**, 251-260.
4.      F. T. Wallenberger, R. J. Hicks and A. T. Bierhals, *J. Non-Cryst. Solids*, 2004, **349**, 377-387.
5.      A. Pönitzsch, M. Nofz, L. Wondraczek and J. Deubener, *J. Non-Cryst. Solids*, 2016, **434**, 1-12.
6.      N. Jakse, M. Bouhadja, J. Kozaily, J. Drewitt, L. Hennet, D. Neuville, H. Fischer, V. Cristiglio and A. Pasturel, *Am. Mineral.*, 2012, **101**, 201903.
7.      M. Bouhadja, N. Jakse and A. Pasturel, *J. Chem. Phys.*, 2014, **140**, 234507.
8.      B. Bunker, *J. Non-Cryst. Solids*, 1994, **179**, 300-308.
9.      T. K. Bechgaard, A. Goel, R. E. Youngman, J. C. Mauro, S. J. Rzoska, M. Bockowski, L. R. Jensen and M. M. Smedskjaer, *J. Non-Cryst. Solids*, 2016, **441**, 49-57.
10.     L. Hennet, J. W. Drewitt, D. R. Neuville, V. Cristiglio, J. Kozaily, S. Brassamin, D. Zanghi and H. E. Fischer, *J. Non-Cryst. Solids*, 2016, **451**, 89-93.
11.     D. R. Neuville, L. Cormier, V. Montouillout and D. Massiot, *J. Non-Cryst. Solids*, 2007, **353**, 180-184.
12.     D. R. Neuville, L. Cormier and D. Massiot, *Chem. Geol.*, 2006, **229**, 173-185.
13.     S. K. Lee and J. F. Stebbins, *Geochim. Cosmochim. Ac.*, 2006, **70**, 4275-4286.
14.     L. Cormier, D. R. Neuville and G. Calas, *J. Am. Ceram. Soc.*, 2005, **88**, 2292-2299.
15.     D. R. Neuville, L. Cormier and D. Massiot, *Geochim. Cosmochim. Ac.*, 2004, **68**, 5071-5079.
16.     D. R. Neuville, L. Cormier, A.-M. Flank, V. Briois and D. Massiot, *Chem. Geol.*, 2004, **213**, 153-163.
17.     J. R. Allwardt, S. K. Lee and J. F. Stebbins, *Am. Mineral.*, 2003, **88**, 949-954.
18.     V. Petkov, S. J. L. Billinge, S. D. Shastri and B. Himmel, *Phys. Rev. Lett.*, 2000, **85**, 3436.
19.     L. Cormier, D. R. Neuville and G. Calas, *J. Non-Cryst. Solids*, 2000, **274**, 110-114.
20.     S. K. Lee and J. F. Stebbins, *Am. Mineral.*, 1999, **84**, 937-945.

21.    J. F. Stebbins and Z. Xu, *Nature*, 1997, **390**, 60.
22.    S. Takahashi, D. R. Neuville and H. Takebe, *J. Non-Cryst. Solids*, 2015, **411**, 5-12.
23.    S. Y. Park and S. K. Lee, *Geochim. Cosmochim. Ac.*, 2012, **80**, 125-142.
24.    M. Moesgaard, R. Keding, J. Skibsted and Y. Yue, *Chem. Mater.*, 2010, **22**, 4471-4483.
25.    J. F. Stebbins, E. V. Dubinsky, K. Kanehashi and K. E. Kelsey, *Geochim. Cosmochim. Ac.*, 2008, **72**, 910-925.
26.    P. Ganster, M. Benoit, J.-M. Delaye and W. Kob, *Mol. Simulat.*, 2007, **33**, 1093-1103.
27.    P. Ganster, M. Benoit, W. Kob and J.-M. Delaye, *J. Chem. Phys.*, 2004, **120**, 10172-10181.
28.    L. Cormier, D. Ghaleb, D. R. Neuville, J.-M. Delaye and G. Calas, *J. Non-Cryst. Solids*, 2003, **332**, 255-270.
29.    M. Benoit, S. Ispas and M. E. Tuckerman, *Phys. Rev. B*, 2001, **64**, 224205.
30.    H. Jabraoui, M. Badawi, S. Lebègue and Y. Vaills, *J. Non-Cryst. Solids*, 2018, **499**, 142-152.
31.    M. Bauchy, *J. Chem. Phys.*, 2014, **141**, 024507.
32.    M. Bouhadja, N. Jakse and A. Pasturel, *J. Chem. Phys.*, 2013, **138**, 224510.
33.    A. Tandia, N. T. Timofeev, J. C. Mauro and K. D. Vargheese, *J. Non-Cryst. Solids*, 2011, **357**, 1780-1786.
34.    D. R. Neuville, L. Cormier, V. Montouillout, P. Florian, F. Millot, J.-C. Rifflet and D. Massiot, *Am. Mineral.*, 2008, **93**, 1721-1731.
35.    D. R. Neuville, L. Cormier, V. Montouillout, P. Florian, F. Millot, J.-C. Rifflet and D. Massiot, *Am. Mineral.*, 2008, **93**, 1721-1731.
36.    S. K. Lee, G. D. Cody and B. O. Mysen, *Am. Mineral.*, 2005, **90**, 1393-1401.
37.    K. Gong and C. E. White, *Cem. Concr. Res.*, 2016, **89**, 310-319
38.    C. Jiang, K. Li, J. Zhang, Q. Qin, Z. Liu, M. Sun, Z. Wang and W. Liang, *J. Non-Cryst. Solids*, 2018, **502**, 76-82.
39.    M. Ben Haha, B. Lothenbach, G. Le Saout and F. Winnefeld, *Cem. Concr. Res.*, 2011, **41**, 955-963.
40.    M. Ben Haha, B. Lothenbach, G. Le Saout and F. Winnefeld, *Cem. Concr. Res.*, 2012, **42**, 74-83.
41.    S. A. Bernal, R. San Nicolas, R. J. Myers, R. M. de Gutiérrez, F. Puertas, J. S. J. van Deventer and J. L. Provis, *Cem. Concr. Res.*, 2014, **57**, 33-43.
42.    R. Snellings, T. Paulhiac and K. Scrivener, *Waste Biomass Valori.*, 2014, **5**, 369-383.
43.    J. L. Provis and J. S. J. van Deventer, eds., *Alkali activated materials: state-of-the-art report, RILEM TC 224-AAM*, Springer/RILEM, Dordrecht, 2014.
44.    J. Skibsted and R. Snellings, *Cem. Concr. Res.*, 2019, **124**, 105799.
45.    P. J. M. Monteiro, S. A. Miller and A. Horvath, *Nat. Mater.*, 2017, **16**, 698-699.
46.    S. Krämer, J. Yang, C. G. Levi and C. A. Johnson, *J. Am. Ceram. Soc.*, 2006, **89**, 3167-3175.
47.    D. L. Poerschke, R. W. Jackson and C. G. Levi, *Ann. Rev. Mater. Res.*, 2017, **47**, 297-330.
48.    T. Oey, A. Kumar, I. Pignatelli, Y. Yu, N. Neithalath, J. W. Bullard, M. Bauchy and G. Sant, *J. Am. Ceram. Soc.*, 2017, **100**, 5521-5527.
49.    R. Snellings, *J. Am. Ceram. Soc.*, 2015, **98**, 303-314.
50.    X. Ke, S. A. Bernal and J. L. Provis, *Cem. Concr. Res.*, 2016, **81**, 24-37.
51.    P. Wang, R. Trettin and V. Rudert, *Adv. Cem. Res.*, 2005, **17**, 161-167.
52.    E. Douglas and J. Brandstetr, *Cem. Concr. Res.*, 1990, **20**, 746-756.
53.    A. E. Morandeau and C. E. White, *Chem. Mater.*, 2015, **27**, 6625-6634.

54.     L. Cormier and D. R. Neuville, *Chem. Geol.*, 2004, **213**, 103-113.
55.     C. Jiang, K. Li, J. Zhang, Q. Qin, Z. Liu, W. Liang, M. Sun and Z. Wang, *J. Mol. Liq.*, 2018, **268**, 762-769.
56.     C. Jiang, K. Li, J. Zhang, Q. Qin, Z. Liu, W. Liang, M. Sun and Z. Wang, *Metall. Mater. Trans. B*, 2018, 1-9.
57.     D. Liang, Z. Yan, X. Lv, J. Zhang and C. Bai, *Metall. Mater. Trans. B*, 2017, **48**, 573-581.
58.     L. Mongalo, A. S. Lopis and G. A. Venter, *J. Non-Cryst. Solids*, 2016, **452**, 194-202.
59.     K. Shimoda and K. Saito, *ISIJ Int.*, 2007, **47**, 1275-1279.
60.     M. González, *École thématique de la Société Française de la Neutronique*, 2011, **12**, 169-200.
61.     R. Vuilleumier, N. Sator and B. Guillot, *Geochim. Cosmochim. Ac.*, 2009, **73**, 6313-6339.
62.     L. Deng and J. Du, *J. Chem. Phys.*, 2018, **148**, 024504.
63.     X. Li, W. Song, K. Yang, N. A. Krishnan, B. Wang, M. M. Smedskjaer, J. C. Mauro, G. Sant, M. Balonis and M. Bauchy, *J. Chem. Phys.*, 2017, **147**, 074501.
64.     P. J. Chupas, K. W. Chapman and P. L. Lee, *J. Appl. Crystallogr.*, 2007, **40**, 463-470.
65.     A. P. Hammersley, S. O. Svensson, M. Hanfland, A. N. Fitch and D. Hausermann, *Int. J. High Pressure Res.*, 1996, **14**, 235-248.
66.     A. P. Hammersley, *FIT2D: An introduction and overview. European synchrotron radiation facility internal report* Report ESRF97HA02T, European Synchrotron Radiation Facility, Grenoble, France, 1997.
67.     T. Egami and S. J. L. Billinge, *Underneath the Bragg peaks: Structural analysis of complex materials*, Pergamon, Elmsford, NY, 2003.
68.     X. Qiu, J. W. Thompson and S. J. L. Billinge, *J. Appl. Crystallogr.*, 2004, **37**, 678-678.
69.     C. L. Farrow, P. Juhas, J. W. Liu, D. Bryndin, E. S. Božin, J. Bloch, T. Proffen and S. J. L. Billinge, *J. Phys.: Condens. Matter*, 2007, **19**, 335219.
70.     T. Proffen, T. Egami, S. Billinge, A. Cheetham, D. Louca and J. Parise, *Appl. Phys. A-Mater.*, 2002, **74**, s163-s165.
71.     P. F. Peterson, M. Gutmann, T. Proffen and S. J. L. Billinge, *J. Appl. Crystallogr.*, 2000, **33**, 1192-1192.
72.     K. Page, C. E. White, E. G. Estell, R. B. Neder, A. Llobet and T. Proffen, *J. Appl. Crystallogr.*, 2011, **44**, 532-539.
73.     C. E. White, N. J. Henson, L. L. Daemen, M. Hartl and K. Page, *Chem. Mater.*, 2014, **26**, 2693-2702.
74.     C. E. White, J. L. Provis, T. Proffen, D. P. Riley and J. S. J. van Deventer, *Phys. Chem. Chem. Phys.*, 2010, **12**, 3239-3245.
75.     P. Biswas, R. Atta-Fynn and D. A. Drabold, *Phys. Rev. B*, 2007, **76**, 125210.
76.     R. Atta-Fynn, D. A. Drabold, S. R. Elliott and P. Biswas, *Phys. Rev. Mater.*, 2018, **2**, 115602.
77.     QuantumATK version 2016.4, QuantumWise A/S).
78.     J. Schneider, J. Hamaekers, S. T. Chill, S. Smidstrup, J. Bulin, R. Thesen, A. Blom and K. Stokbro, *Model. Simul. Mater. Sc.*, 2017, **25**, 085007.
79.     K. C. Mills, L. Yuan and R. T. Jones, *J. S. Afr. I. Min. Metall.*, 2011, **111**, 649-658.
80.     M. Matsui, *Mineral. Mag.*, 1994, **58**, 571-572.
81.     G. Kresse and J. Furthmüller, *Phys. Rev. B*, 1996, **54**, 11169.
82.     M. G. Tucker, D. A. Keen, M. T. Dove, A. L. Goodwin and Q. Hui, *J. Phys.: Condens. Matter*, 2007, **19**, 335218.

83.  R. N. Mead and G. Mountjoy, *J. Phys. Chem. B*, 2006, **110**, 14273-14278.
84.  P. Gaskell, M. Eckersley, A. Barnes and P. Chieux, *Nature*, 1991, **350**, 675.
85.  M. Guignard and L. Cormier, *Chem. Geol.*, 2008, **256**, 111-118.
86.  K. Vollmayr, W. Kob and K. Binder, *Phys. Rev. B*, 1996, **54**, 15808.
87.  C. Hühn, L. Wondraczek and M. Sierka, *Phys. Chem. Chem. Phys.*, 2015, **17**, 27488-27495.
88.  K. Shimoda, Y. Tobu, K. Kanehashi, T. Nemoto and K. Saito, *J. Non-Cryst. Solids*, 2008, **354**, 1036-1043.
89.  M. Ren, J. Y. Cheng, S. P. Jaccani, S. Kapoor, R. E. Youngman, L. Huang, J. Du and A. Goel, *J. Non-Cryst. Solids*, 2019, **505**, 144-153.
90.  M. J. Toplis, S. C. Kohn, M. E. Smith and I. J. Poplett, *Am. Mineral.*, 2000, **85**, 1556-1560.
91.  S. K. Lee, H.-I. Kim, E. J. Kim, K. Y. Mun and S. Ryu, *J. Phys. Chem. C*, 2015, **120**, 737-749.
92.  N. Trcera, D. Cabaret, S. Rossano, F. Farges, A.-M. Flank and P. Lagarde, *Phys. Chem. Miner.*, 2009, **36**, 241-257.
93.  L. Cormier and G. J. Cuello, *Phys. Rev. B*, 2011, **83**, 224204.
94.  K. Shimoda, Y. Tobu, M. Hatakeyama, T. Nemoto and K. Saito, *Am. Mineral.*, 2007, **92**, 695-698.
95.  S. Kroeker and J. F. Stebbins, *Am. Mineral.*, 2000, **85**, 1459-1464.
96.  A. Navrotsky, H. D. Zimmermann and R. L. Hervig, *Geochim. Cosmochim. Ac.*, 1983, **47**, 1535-1538.
97.  S. K. Lee and S. Sung, *Chem. Geol.*, 2008, **256**, 326-333.
98.  J. R. Allwardt and J. F. Stebbins, *Am. Mineral.*, 2004, **89**, 777-784.
99.  G. S. Frankel, J. D. Vienna, J. Lian, J. R. Scully, S. Gin, J. V. Ryan, J. Wang, S. H. Kim, W. Windl and J. Du, *NPJ Mater. Degrad.*, 2018, **2**, 15.
100.  J. Du and J. M. Rimsza, *NPJ Mater. Degrad.*, 2017, **1**, 16.